# Coexistence of Superconductivity and Chern insulators with $|C| \geq 1$ in twisted rhombohedral bilayer-trilayer graphene.

Qiu Yang[1,2*], Zihao Huo[1,2*†], Le Liu[3*], Tianqi Song[2,4], Fankun Wang[2,4], Jingwei Dong[2,4], Kenji Watanabe[5], Takashi Taniguchi[6], Weibo Gao[3], Xiaobo Lu[1,7], Guangyu Zhang[2,4], Wei Yang[2,4†]

[1] *International Center for Quantum Materials, School of Physics, Peking University, Beijing 100871, China*

[2] *Beijing National Laboratory for Condensed Matter Physics and Institute of Physics, Chinese Academy of Sciences, Beijing 100190, China*

[3] *School of Electrical and Electronic Engineering, Nanyang Technological University, Singapore 639798, Singapore*

[4] *School of Physical Sciences, University of Chinese Academy of Sciences, Beijing, 100190, China*

[5] *Research Center for Electronic and Optical Materials, National Institute for Material Science, 1-1 Namiki, Tsukuba 305-0044, Japan.*

[6] *Research Center for Materials Nanoarchitectonics, National Institute for Material Science, 1-1 Namiki, Tsukuba 305-0044, Japan*

[7] *Interdisciplinary Institute of Light-Element Quantum Materials and Research Centre for Light-Element Advanced Materials, Peking University, Beijing 100407, China.*

*These authors contributed equally to this work.

† Corresponding authors. Emails: zhhuo@stu.pku.edu.cn; wei.yang@iphy.ac.cn

**Abstract:** Rhombohedral multilayer graphene (RMG) has been demonstrated as a promising platform for investigating strongly correlated phases of matter. Here, we observe a coexistence of superconductivity and Chern insulators in a twisted rhombohedral bilayer-trilayer graphene moiré superlattice. Superconductivity, evidenced by a sharp transition with a critical temperature of ~180 mK, nonlinear *I-V* with a BKT temperature of 135 mK, and a signature of phase-sensitive interference patterns, emerges from a symmetry-breaking Fermi surface near $\nu = -2$ and $D = 0$, and it survives at a large in-plane magnetic field up to 2 T, strongly deviating from the Pauli limit. Moreover, we observe a strong modulation of the Ginzburg-Landau coherence length from 467 nm to 145 nm as the in-plane magnetic field increases from 0 to 2 T, whereas the critical current $I_c$ barely changes. Unlike previous studies in RMG, we observe a competition between superconductivity and a correlated insulator at $\nu = -2$, monitored by displacement fields and twist angles. Note that we also observe integer Chern insulators with $C = 1$, 3, and -3 at $\nu = 1/2$, 1, and 3, respectively, for large $D > 0$ in the same superconducting device. The observations establish twisted rhombohedral graphene multilayers as a fascinating system to investigate unconventional superconductivity, high Chern number topology, and their interplay.

**Main text**

Rhombohedral multilayer graphene has been demonstrated as a promising platform for investigating strongly correlated phases of matter and unconventional superconductivity (SC) [1–12], owing to its unique low-energy flatband structure [13–15]. Recently, superconductivity has been widely observed in RMG from trilayer to more than 10 layers [4,9,11,16–23], exhibiting unconventional magnetic field properties, e.g., strong deviations from the Pauli limit or even magnetic-field-boosted SC [4,16–19,23,24]. Even more strikingly, experimental evidence of chiral superconductivity with anomalous Hall effects has been observed in R4G, R5G, R6G at high displacement fields [11,25], showing great promise for realizing topological superconductivity. Note that for all these studies, superconductivity is intimately related to high density of states (DOS) and symmetry-breaking phase boundaries, at which intertwined orders and quantum fluctuations of isospins might play critical roles.

RMG also provides fertile ground for topological phases due to its nontrivial Berry curvature. By tuning the flatness of the band dispersion via electric displacement fields, the enhanced DOS gives rise to Fermi surface instability and various spontaneous isospin polarizations, including magnetism [3,5,26,27]. By further introducing moiré modulation via hBN alignment, the boosted electron correlation creates an isolated topological flatband with Chern number $C = 1$ at moiré filling $\nu = 1$ [28,29], leading to the discovery of integer and fractional Chern insulators at zero magnetic field [8,10,12,30–32], providing lattice analogues of quantum Hall states in Landau levels. Recently, a new strategy by twisting RMG has been developed for band topology engineering, leading to isolated moiré flatbands with high Chern numbers ($|C| > 1$), as confirmed by calculations [33] and quantum anomalous Hall (QAH) effects [34–38]. Importantly, signatures of fractionalization in high Chern number flat bands have been reported in twisted rhombohedral bilayer-trilayer [34] and bilayer-tetralayer graphene [38], raising the prospect of fractional topology beyond the conventional Landau level paradigm.

Here, we observe a coexistence of SC and Chern insulators with $|C| \geq 1$ in twisted rhombohedral bilayer-trilayer graphene moiré superlattice, aiming to bridge the gap between SC and topology. Previous work [34] lies in the optimal twist angle for electron correlation, with twist angle $\theta$ ranging from 1.3° to 1.5°, where correlated insulators were observed at $\nu = -2$ on the moiré valence band near $D = 0$, and more prominently, Chern insulators were discovered between $\nu = 1$ and 3 on the moiré conduction flatband for $D > 0$. Now we focus on a smaller angle with $\theta = 1.20°$ to reduce the interaction and tip the competing orders at $\nu = -2$ near $D = 0$, while retaining strong correlation in the much flatter moiré conduction bands for $D > 0$. We observe a SC pocket near $\nu = -2$ and $D = 0$, competing against a correlated insulator at $D < 0$, and we identify QAH states emerging at $\nu = 1/2$, 1, and 3, with Chern numbers $C = +1$, $+3$, and $-3$, respectively.

***Stacking-dependent transport in twisted bilayer-trilayer graphene.*** The devices consist of a Bernal-stacked bilayer graphene twisted relative to a Bernal (ABA) or rhombohedral (ABC) stacked trilayer, denoted 2+ABA and 2+ABC configurations, respectively [Fig. 1(a) and Fig. S1, see also Methods in Supplementary Materials]. Top and Bottom graphite gates independently tune the carrier density $n$ and displacement field $D$. Figures 1(b) and 1(c) show the longitudinal resistance $R_{xx}$ at $T = 1.5$ K for D1 (2+ABC, $\theta = 1.20°$) and D2 (2+ABA, $\theta = 1.19°$), respectively, with a logarithmic color scale in Fig. 1(c). Despite their similar twist angles, the devices exhibit markedly different resistance landscapes. In particular, D1 displays pronounced correlated insulating behavior near integer fillings on the electron-doped side, whereas these behaviors are much weaker in D2. Similar correlated insulating behaviors are also observed at $T = 1.5$ K in three additional 2+ABC devices, D3-D5, with different twist angles from 1.21° to 1.35° [Fig. S2]. The contrast between 2+ABA and 2+ABC is further illustrated by the linecuts at $V_{tg} = 3.25$ V in Fig. 1(d), taken along the dashed trajectories in Figs. 1(b) and 1(c), with $R_{xx}$ of 2+ABA multiplied by ten for comparison.

Band-structure calculations using a non-interacting continuum model provide further insight into the contrasting behaviour between 2+ABA and 2+ABC. At full filling, $\nu = +4$, the two stacking configurations exhibit markedly different displacement-field dependences [Fig. 1(e)]. In 2+ABC, the resistance is maximal near $D = 0$ and decreases as $D$ increases. While in 2+ABA, it exhibits high resistance at negative $D$, followed by low resistance at small positive $D$ and a renewed increase at larger positive $D$. The calculated gaps qualitatively reproduce this contrast, establishing a band-structure basis for the observed transport behaviour. The calculated DOS across the $\nu$-$D$ plane further captures the overall trends of the transport phase diagrams [Fig. 1(f), (g)]. In particular, 2+ABC exhibits an enhanced DOS on the electron-doped side at positive $D$, where wavefunctions are polarized towards the trilayer [Fig. S3(f)], coinciding with the correlated insulating states in experiments. The corresponding DOS enhancement is nearly absent in 2+ABA. These observations agree well with the previous reports with larger twist angles [34,39], demonstrating the crucial role played by rhombohedral stacking. In the following, we focus mainly on D1 to examine superconductivity and QAH states in the 2+ABC configuration.

***Superconducting pocket near $\nu = -2$.*** Figure 2(a) shows a comparison of the phase diagram near $\nu = -2$ and $D = 0$ between a base temperature of 10 mK and an elevated temperature of 300 mK. We observe evidence of a SC pocket near $\nu = -2$ for a small range of $D/\varepsilon_0$ from 0.05 to -0.04 V/nm, indicated by the dark blue color, and a correlated insulator in proximity at $\nu = -2$ for $D/\varepsilon_0 < -0.04$ V/nm, highlighted by the red color. Figure 2(b) shows typical temperature dependence of transfer curves along a line cut through the SC pocket. The $R$-$T$ curves exhibit sharp transitions from SC to metallic behaviour as $T$ increases [inset of Fig. 2(b)], with a SC critical temperature maximized around 180 mK. Note that there is a small finite resistance of approximately 20 Ω in the SC state. The SC transition exhibits a dome-shaped profile across the pocket [Fig. 2(c)], as determined from the $T$-dependent resistance traces using the criterion $R_{xx}(T_c) = 0.5\, R_N$, where $R_N$ is the normal-state resistance.

In addition, characteristic nonlinear differential resistance $dV_{xx}/dI$ curves against d.c. current $I_{dc}$ provides further evidence for SC [inset of Fig. 2(d)]. We observe a Berezinskii–Kosterlitz–Thouless (BKT) transition temperature $T_{BKT} \approx 135$ mK [Fig. 2(d)], identified by $V_{xx} \propto I_{dc}^3$ (equivalent to $dV_{xx}/dI \propto I_{dc}^2$) [40,41], and its recovery to linear metallic state at $T > 250$ mK. The SC state is further demonstrated by its extreme sensitivity to the out-of-plane magnetic fields ($B_\perp$), with a critical $B_\perp$ on the millitesla scale and characteristic spindle-shaped features with a Fraunhofer-like pattern [Fig. 2(e)]. At $D = 0$, the differential resistance $dV_{xx}$/dI exhibits a filling-dependent nonlinear response to $I_{dc}$ [Fig. 2(f)], where the critical current shares a dome structure that resembles the filling dependence of $T_c$. Additional measurements of the perpendicular-field response at different fillings and displacement fields are shown in Fig. S4. A similar SC pocket was also observed in device D3 (2+ABC with $\theta = 1.21°$ in Fig. S11), confirming its reproducibility.

***Fermiology of the SC pocket.*** The phase structure surrounding the superconducting pocket suggests a close connection between superconductivity and spin polarization. The SC pocket borders a highly resistive region at more negative displacement fields [Fig. 2(a)], whose transport behaviors are consistent with a spin-polarized correlated insulator. By applying an in-plane magnetic field $B_{//} = 2$ T, the insulator region expands, and by contrast, the SC pocket shrinks in phase space [Figs. S5(e), (f)]. Notably, a lobe-shaped metallic phase expands between the SC and insulating states [Fig. 3(a); see also Figs. S5(a), (d)]. Sweeping either the top- or bottom-gate voltage across the boundary produces hysteresis in $R_{xx}$ [Figs. 3(b), (c)], suggesting a first-order phase transition [3,42]. We investigate the states separated by this boundary through quantum oscillations measured at $B_{//}$ = 2 T and $B_{\perp}$ = 1 T [Fig. 3(d) and Methods], supported by detailed Landau fan analysis [Fig. S8 and Methods]. Inside the metallic region, the oscillations extracted after background subtraction indicate a one-fold degeneracy, g ≈ 1 [Fig. 3(e), see also Fig. S7 and Methods], and the Landau fan trajectories remain traceable over a broader perpendicular-field range [Fig. S8(b)]. These observations favour a quarter-metal (QM) interpretation of the enclosed phase. In low-doping regions outside the enclosed region, the fan trajectories are interrupted over relatively short field ranges [Fig. S8(c)], and the frequency peak at $f_\nu = 1/4$ [Fig. S8(d) and S8(f)] supports a four-fold-degenerate normal metal. Temperature-dependent measurements along the trajectory marked in Fig. S5(g) further distinguish different regimes [Fig. S5(h)].

The SC shows unconventional in-plane magnetic field dependence. The characteristic SC evidence persists in in-plane magnetic fields up to 2T, which indicates a strong Pauli limit deviation [See Methods]. More strikingly, we observe a substantial enhancement of the perpendicular critical field $B_{\perp c}$ [Figs. 2(e) and S5(b), (c)] as $B_{//}$ increases. To quantify this evolution, we analyze $B_{\perp c}(T)$ using Ginzburg–Landau (GL) theory, $B_{\perp c} = (\Phi_0 / (2\pi\xi^2))(1 - T/T_c)$, where $\Phi_0$ is the superconducting flux quantum. The linear fit of $B_{\perp c}(T)$ yields a coherence length that approximates the SC coupling strength. Profoundly, we observe a strong modulation of the coherence length by in-plane magnetic fields in Fig. 3(f) [see also Fig. S6], in which $\xi_{GL}$ decreases from 467 nm at $B_{//} = 0$ to 145 nm at $B_{//}$ = 2 T. The observations conclusively indicate more tightly bound Cooper pairs under in-plane fields, and motivate a scenario of a spin-polarized superconductivity.

***Quantum anomalous Hall states on the electron-doped side.*** Aside from the SC near $\nu = -2$ and $D = 0$, we observe a proliferation of Chern insulators in the topological moiré conduction band under large positive $D$ in the same device D1($\theta$ = 1.20°). Figures 4(a) and 4(b) show maps of symmetrized longitudinal resistivity $\rho_{xx}$ and antisymmetrized Hall resistivity $\rho_{xy}$ against $\nu$ and $D$ obtained from measurements at $B_{\perp} = \pm 0.2$ T. The nature of the topological phases is revealed from the chiral edge states and the bulk insulating states, yielding vanishing $\rho_{xx}$ and quantized $\rho_{xy} = h/(Ce^2)$ for a Chern insulator. We first confirm the high Chern number of the moiré conduction band from the observation of a Chern insulator with $C = 3$ at $\nu = 1$, corresponding to the $\rho_{xx}$ dip and $\rho_{xy}$ plateau at $\nu = 1$ [Fig. 4(a)]. The Chern insulator is further verified from the Streda analysis in the Landau fan diagram according to $\partial n/\partial B_{\perp} = C\cdot e/h$ [Fig. 4(c)], and more directly from the QAH effect with magnetic hysteresis [Fig. 4(e)]. In addition, we also identify a symmetry-breaking Chern insulator with $C = 1$ at $\nu = 1/2$ [Fig. 4(a)], whose topological nature is further revealed in Figs. 4(c) and 4(e). Similarly, we also observe a Chern insulator with $C = -3$ at $\nu = 3$ [Fig. 4(b)], verified from Chern number analysis [Fig. 4(d)] and QAH effect [Fig. 4(e)]. Similar Chern insulating states are also observed in devices D3-D5. The observations of high Chern insulators at integer fillings $\nu = 1$ and 3 agree well with previous work with twist angles ranging from 1.3 to 1.5 degrees [34], demonstrating the robustness of the high Chern number flatband. Note that the topological phases at fractional fillings are more sensitive to the twist angle, calling for additional work for further exploration.

***Discussions and conclusions.*** Having established the coexistence of superconductivity and Chern insulators in the same device, we further highlight a competition between superconductivity and a correlated insulator. As the twist angle $\theta$ increases from 1.20° to 1.31°, a correlated insulator develops near $\nu = -2$ and $D = 0$ [Fig. S12(a)]. The comparison is more visible in the resistance traces at $D = 0$ [Fig. S12(b)]. Both the SC and the correlated insulator occupy bounded regions near $\nu$ = -2, with sharp resistance jumps at their boundaries, suggesting a shared tendency towards spontaneous symmetry-breaking Fermi surface. To distinguish the reconstruction and polarization of the Fermi surface, we analyze the relative Hall filling $\nu_H$-$\nu_0$ across different ranges of $B_\perp$ [Figs.3(g)-(i); see also Fig. S10], whose nonzero integers indicate new polarized Fermi surfaces while divergences mark the symmetry-breaking boundary. At $|B_\perp| < 0.1$ T and $B_{//} = 0$, $\nu_H$-$\nu_0$ in the SC pocket is around 3.5 [Fig. 3(h)], indicating a partially polarized Fermi surface that becomes further polarized with increasing $B_{//}$. The low-field Landau fan diagram in the SC regime additionally indicates coexistence of electron and hole pockets, supporting a partially polarized Fermi surface with band inversion [Figs. S8(e), S9(a)]. By contrast, the correlated insulator regime in D5 ($\theta = 1.31°$) at $B_{//} = 0$ exhibits $\nu_H$-$\nu_0 \approx 2$ [Fig. S12(c)], suggesting a polarized Fermi surface developed from $\nu$ = -2. Interestingly, by applying a larger out-of-plane magnetic field up to 0.6 T, the SC pocket evolves into an incipient correlated insulator with a polarized Fermi surface with $\nu_H$-$\nu_0 \approx 2$[Figs. 3(g), (i)], recovering the same features in D5. These observations offer a new perspective on the nature of SC and its competition with a correlated insulator, monitored by twist angle.

In moiré flatbands, the characteristic Coulomb energy scales approximately as $U \propto 1/(\varepsilon_r \lambda_M)$, where $\varepsilon_r$ is the effective dielectric constant and $\lambda_M$ is the moiré period. Note that D1 with 1.20° and D5 with 1.31° share the same bilayer graphene and dielectric hBN stacks, providing a comparable dielectric environment. Presumably, the shorter $\lambda_M$ at larger $\theta$ therefore suggests stronger Coulomb interaction. By further approximating the kinetic energy from the moiré valence band bandwidth, the $\theta$-dependent competition between SC and correlated insulator is analogous to competing orders driven by $U/t$. This behavior closely resembles SC in twisted $WSe_2$ [43,44], and high $T_c$ superconductors [45,46], where moderate $U/t$ favors SC while large $U/t$ prefers the correlated insulator. Note that such a competition can also be tuned by out-of-plane magnetic fields or displacement fields in addition to twist angle control.

To sum up, we observe superconductivity and Chern insulators with $|C| \geq 1$ in twisted rhombohedral bilayer-trilayer graphene near 1.20°. A superconducting pocket near $\nu = -2$ reaches $T_c \approx 180$ mK and exhibits BKT transition temperature $T_{BKT} \approx 135$ mK. The SC violates the conventional Pauli limit, and in-plane magnetic fields further enhance the perpendicular critical field. Twist angle and displacement field shift the balance between SC and correlated insulator, implicating Fermi-surface reconstruction and isospin polarization in the competition. On the electron-doped side, QAH states appear at $\nu$ = 1/2, 1, and 3, with Chern numbers $C$ = +1, +3, and -3, respectively. Our work establishes twisted rhombohedral bilayer-trilayer graphene as a tunable platform for investigating unconventional superconductivity, topology, and their interplay, showing great promise for novel quantum devices.

***Note added:*** During the preparation of this paper, we became aware of a related study on similar states in twisted bilayer-trilayer graphene [47].

***Acknowledgements.*** W.Y. acknowledges support from the Quantum Science and Technology-National Science and Technology Major Project (Grant No. 2025ZD0300500), the Strategic Priority Research Program of Chinese Academy of Sciences (No. XDB1710000), the Beijing Natural Science Foundation (JQ25003), and the National Key Research and Development Program (Grant No. 2024YFA1410400). X.L. discloses support for the research of this work from the National Key R&D Program (Grant No. 2022YFA1403502 and 2024YFA1409002), the National Natural Science Foundation of China (Grant No. 12521006 and 12274006), and the Beijing National Laboratory for Condensed Matter Physics (Grant No. 2025BNLCMPKF001). K.W. and T.T. acknowledge support from the Elemental Strategy Initiative conducted by the MEXT, Japan (Grant No. JPMXP0112101001), JSPS KAKENHI (Grant Nos. 19H05790, 20H00354 and 21H05233) and A3 Foresight by JSPS.

***Author Contributions.*** W.Y. and X.L. conceived and designed the experiments; Q.Y., Z.H. fabricated the devices and performed the transport measurements with help from T.S., F.W. and J.D.; L.L. performed the theoretical modeling; T.T. and K.W. contributed hBN substrates; Q.Y., Z.H., L.L., W.G., X.L., G.Z. and W.Y. analyzed the data; Q.Y., Z.H. and W.Y. wrote the paper with input from others.

The authors declare no competing interests.

***Data and Code Availability.*** All data supporting the findings of this study are available within the main text, figures and Supplementary Materials, or from the corresponding authors upon request. Source data are provided with this paper. Codes that support the findings of this study are available upon request. Codes include scripts for data processing and theoretical modelling.

***References***

[1] G. Chen et al., Evidence of a gate-tunable Mott insulator in a trilayer graphene moiré superlattice, Nat. Phys. **15**, 237 (2019).
[2] G. Chen et al., Tunable correlated Chern insulator and ferromagnetism in a moiré superlattice, Nature **579**, 56 (2020).
[3] H. Zhou et al., Half- and quarter-metals in rhombohedral trilayer graphene, Nature **598**, 429 (2021).
[4] H. Zhou, T. Xie, T. Taniguchi, K. Watanabe, and A. F. Young, Superconductivity in rhombohedral trilayer graphene, Nature **598**, 434 (2021).
[5] T. Han et al., Orbital multiferroicity in pentalayer rhombohedral graphene, Nature **623**, 41 (2023).
[6] T. Han, Z. Lu, G. Scuri, J. Sung, J. Wang, T. Han, K. Watanabe, T. Taniguchi, H. Park, and L. Ju, Correlated insulator and Chern insulators in pentalayer rhombohedral-stacked graphene, Nat. Nanotechnol. **19**, 181 (2024).
[7] K. Liu et al., Spontaneous broken-symmetry insulator and metals in tetralayer rhombohedral graphene, Nat. Nanotechnol. **19**, 188 (2024).
[8] Z. Lu, T. Han, Y. Yao, A. P. Reddy, J. Yang, J. Seo, K. Watanabe, T. Taniguchi, L. Fu, and L. Ju, Fractional quantum anomalous Hall effect in multilayer graphene, Nature **626**, 759 (2024).
[9] Y. Choi et al., Superconductivity and quantized anomalous Hall effect in rhombohedral graphene, Nature **639**, 342 (2025).
[10] Z. Lu et al., Extended quantum anomalous Hall states in graphene/hBN moiré superlattices, Nature **637**, 1090 (2025).
[11] T. Han et al., Signatures of Chiral Superconductivity in Rhombohedral Graphene, Nature **643**, 654 (2025).
[12] J. Xie et al., Tunable fractional Chern insulators in rhombohedral graphene superlattices, Nat. Mater. **24**, 1042 (2025).
[13] H. Min and A. H. MacDonald, Chiral decomposition in the electronic structure of graphene multilayers, Phys. Rev. B **77**, 155416 (2008).
[14] M. Koshino and E. McCann, Trigonal warping and Berry's phase N $\pi$ in ABC-stacked multilayer graphene, Phys. Rev. B **80**, 165409 (2009).
[15] F. Zhang, B. Sahu, H. Min, and A. H. MacDonald, Band structure of ABC-stacked graphene trilayers, Physical Review B **82**, 035409 (2010).

[16]M. Kumar, D. Waleffe, A. Okounkova, R. Tejani, K. Watanabe, T. Taniguchi, É. Lantagne-Hurtubise, J. Folk, and M. Yankowitz, Pervasive Spin-Triplet Superconductivity in Rhombohedral Graphene, arXiv.2511.16578.
[17]Y. Guo et al., Flat Band Surface State Superconductivity in Thick Rhombohedral Graphene, arXiv.2511.17423.
[18]J. Seo et al., Family of magnetic field-boosted superconductors in rhombohedral graphene, Nature **656**, 60 (2026).
[19]J. Xie, Z. Huo, Z. Chen, Z. Zhang, K. Watanabe, T. Taniguchi, X. Lin, and X. Lu, Magnetic-Field-Driven Insulator-Superconductor Transition in Rhombohedral Graphene, Phys. Rev. Lett. **136**, 176505 (2026).
[20]J. Deng, J. Xie, H. Li, T. Taniguchi, K. Watanabe, J. Shan, K. F. Mak, and X. Liu, Superconductivity and ferroelectric orbital magnetism in semimetallic rhombohedral hexalayer graphene, Nat. Phys. (2026).
[21]C. Zheng et al., Multiple Superconducting Phases in Rhombohedral Heptalayer Graphene, Chinese Phys. Lett. **43**, 060718 (2026).
[22]M. Kumar, D. Waleffe, A. Okounkova, R. Tejani, V. T. Phong, K. Watanabe, T. Taniguchi, C. Lewandowski, J. Folk, and M. Yankowitz, Superconductivity from dual-surface carriers in rhombohedral graphene, Nat. Phys. **22**, 862 (2026).
[23]A. Okounkova, A. Sohm, T. Faehndrich, M. Kumar, D. Waleffe, J. Yan, K. Watanabe, T. Taniguchi, J. Folk, and M. Yankowitz, Anomalous metal and superconducting phases in rhombohedral graphene, Nature **658**, 65 (2026).
[24]Z. Huo et al., Spin-Polarized Superconductivity and High-Chern Insulators in Twisted Rhombohedral Graphene Family, arXiv.2609.12504.
[25]Z. Hua et al., Multi-Knob Switchable Chiral Superconductivity Quartet in Rhombohedral Graphene, arXiv.2607.06520.
[26]N. Auerbach et al., Isospin magnetic texture and intervalley exchange interaction in rhombohedral tetralayer graphene, Nature Physics **21**, 1765 (2025).
[27]K. Krötzsch, A. Herasymchuk, Y. Zhumagulov, A. Magrez, K. Watanabe, T. Taniguchi, S. G. Sharapov, O. V. Yazyev, and M. Banerjee, Magnetoelectric Switching of Magnetic Order in Rhombohedral Graphene, arXiv:2509.24672.
[28]Z. Dong, A. S. Patri, and T. Senthil, Theory of Quantum Anomalous Hall Phases in Pentalayer Rhombohedral Graphene Moiré Structures, Physical Review Letters **133**, 206502 (2024).
[29]Z. Guo, X. Lu, B. Xie, and J. Liu, Fractional Chern insulator states in multilayer graphene moiré superlattices, Phys. Rev. B **110**, 075109 (2024).
[30]C. Li et al., Stacking-Orientation and Twist-Angle Control on Integer and Fractional Chern Insulators in Moiré Rhombohedral Graphene, arXiv.2505.01767.
[31]Z. Huo et al., Does Moire Matter? Critical Moire Dependence with Quantum Fluctuations in Graphene Based Integer and Fractional Chern Insulators, arXiv.2510.15309.
[32]J. Xie et al., Unconventional Orbital Magnetism in Graphene-Based Fractional Chern Insulators, arXiv.2506.01485.
[33]P. J. Ledwith, A. Vishwanath, and E. Khalaf, Family of Ideal Chern Flatbands with Arbitrary Chern Number in Chiral Twisted Graphene Multilayers, Physical Review Letters **128**, 176404 (2022).
[34]J. Dong et al., Observation of Integer and Fractional Chern Insulators in High Chern Number Flatbands, arXiv:2507.09908.
[35]W. Wang et al., Programmable Quantum Anomalous Hall Insulator in Twisted Crystalline Flatbands, Phys. Rev. X **16**, 011015 (2026).
[36]X. Wang, L. A. Benítez, S. Rao, V. T. Phong, W. I. Chu, K. Watanabe, T. Taniguchi, C. Lewandowski, and P. Jarillo-Herrero, High-Chern-number orbital magnetism in twisted rhombohedral graphene, Nat. Mater. (2026).
[37]Z. Chen et al., Layer-engineered quantum anomalous Hall effect in twisted rhombohedral graphene, Nat. Mater. (2026).
[38]Z. Li et al., Fractional high-Chern insulator in twisted rhombohedral graphene, Nature **655**, 892 (2026).

[39]J. Dong et al., Symmetry Breaking and Phase Transitions in Twisted Rhombohedral Trilayer-Bilayer Graphene, Chin. Phys. Lett **43**, 060716 (2026).
[40]T. Michael, Introduction to Superconductivity (Courier Corporation, 1996).
[41]Y. Cao, V. Fatemi, S. Fang, K. Watanabe, T. Taniguchi, E. Kaxiras, and P. Jarillo-Herrero, Unconventional superconductivity in magic-angle graphene superlattices, Nature **556**, 43 (2018).
[42]L. Liu et al., Observation of First-Order Quantum Phase Transitions and Ferromagnetism in Twisted Double Bilayer Graphene, Phys. Rev. X **13**, 031015 (2023).
[43]Y. Xia, Z. Han, J. Zhu, Y. Zhang, P. Knüppel, K. Watanabe, T. Taniguchi, K. F. Mak, and J. Shan, Bandwidth-tuned Mott transition and superconductivity in moiré WSe2, Nature **650**, 585 (2026).
[44]Y. Guo et al., Angle evolution of the superconducting phase diagram in twisted bilayer WSe2, Nature **652**, 622 (2026).
[45]P. A. Lee, N. Nagaosa, and X.-G. Wen, Doping a Mott insulator: Physics of high-temperature superconductivity, Rev. Mod. Phys. **78**, 17 (2006).
[46]B. Keimer, S. A. Kivelson, M. R. Norman, S. Uchida, and J. Zaanen, From quantum matter to high-temperature superconductivity in copper oxides, Nature **518**, 179 (2015).
[47]D. Waleffe, A. Bhattacharyya, M. Kumar, E. Maginnis, A. Okounkova, T. Faehndrich, K. Watanabe, T. Taniguchi, J. Folk, and M. Yankowitz, Incipient Superconductivity and Tunable Chern Insulators in Twisted Bilayer-Trilayer Graphene, arXiv:2609.30202.
[48]Z. Feng, W. Wang, Y. You, Y. Chen, K. Watanabe, T. Taniguchi, C. Liu, K. Liu, and X. Lu, Rapid infrared imaging of rhombohedral graphene, Phys. Rev. Applied **23**, 034012 (2025).
[49]C. H. Lui, Z. Li, Z. Chen, P. V. Klimov, L. E. Brus, and T. F. Heinz, Imaging Stacking Order in Few-Layer Graphene, Nano Lett. **11**, 164 (2011).
[50]C. Cong, T. Yu, K. Sato, J. Shang, R. Saito, G. F. Dresselhaus, and M. S. Dresselhaus, Raman Characterization of ABA- and ABC-Stacked Trilayer Graphene, ACS Nano **5**, 8760 (2011).
[51]Y. Shan, Y. Li, D. Huang, Q. Tong, W. Yao, W.-T. Liu, and S. Wu, Stacking symmetry governed second harmonic generation in graphene trilayers, Sci. Adv. **4**, eaat0074 (2018).
[52]N. Lindvall, A. Kalabukhov, and A. Yurgens, Cleaning graphene using atomic force microscope, Journal of Applied Physics **111**, 064904 (2012).
[53]F. R. Geisenhof et al., Anisotropic Strain-Induced Soliton Movement Changes Stacking Order and Band Structure of Graphene Multilayers: Implications for Charge Transport, ACS Appl. Nano Mater. **2**, 6067 (2019).
[54]F. J. Harris, On the use of windows for harmonic analysis with the discrete Fourier transform, Proceedings of the IEEE **66**, 51 (1978).
[55]R. Bistritzer and A. H. MacDonald, Moiré bands in twisted double-layer graphene, Proc. Natl. Acad. Sci. U.S.A. **108**, 12233 (2011).
[56]M. Koshino, Band structure and topological properties of twisted double bilayer graphene, Phys. Rev. B **99**, 235406 (2019).
[57]Y. Guo et al., Flat Band Surface State Superconductivity in Thick Rhombohedral Graphene, arXiv:2511.17423.
[58]K. Kolář, D. Waters, J. Folk, M. Yankowitz, and C. Lewandowski, Single-gate tracking behavior in flat-band multilayer graphene devices, Phys. Rev. B **113**, 075131 (2026).

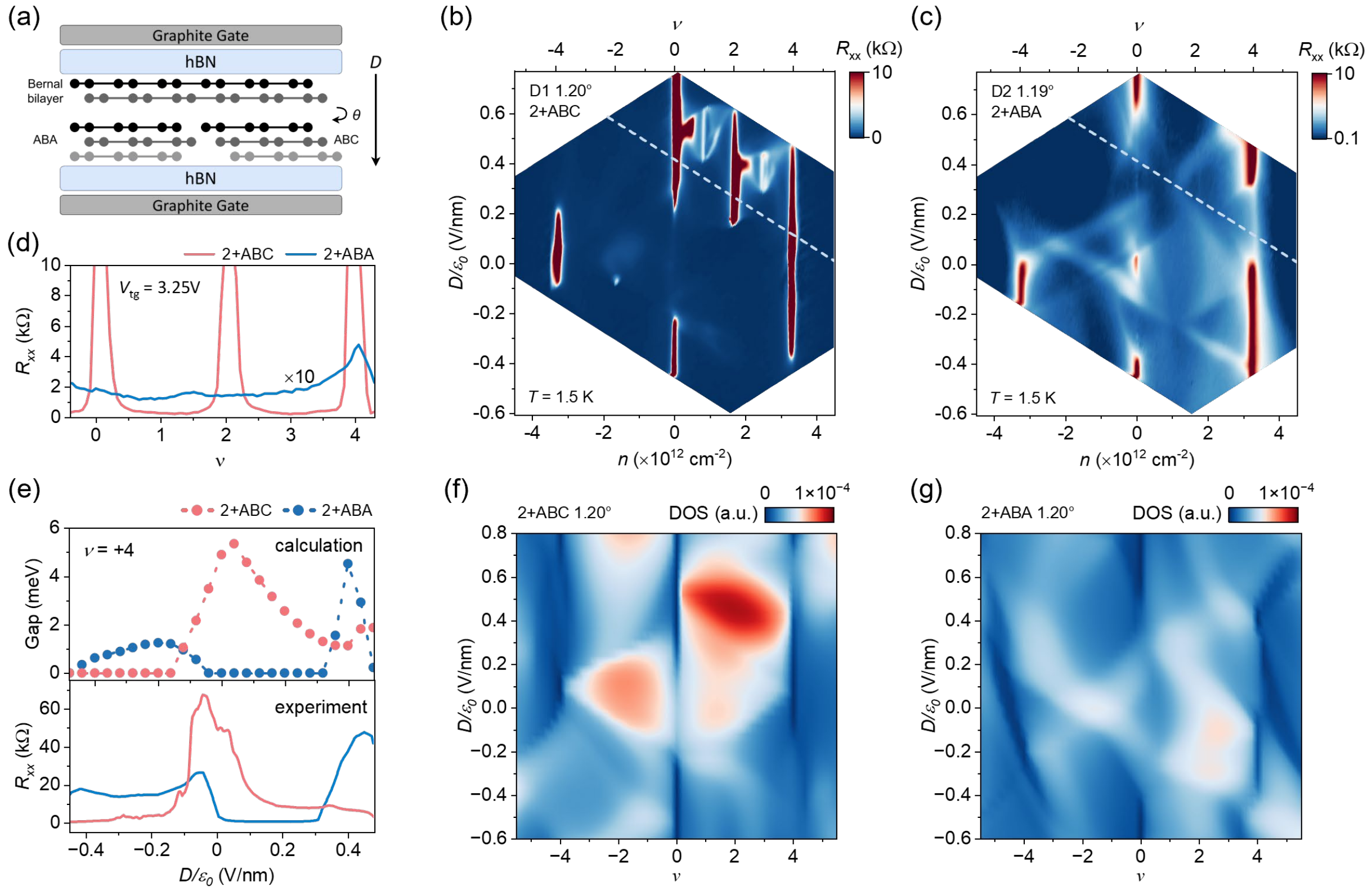

**FIG. 1. Stacking-dependent transport in twisted bilayer-trilayer graphene. (a)** Schematic of the dual-gate device, showing a twisted stack of a Bernal bilayer and an ABA- or ABC-stacked trilayer. These structures are labeled 2+ABA and 2+ABC, respectively. Positive displacement field $D$ points from the top gate toward the bottom gate. **(b), (c)** Maps of the longitudinal resistance $R_{xx}$ versus carrier density $n$ and $D/\varepsilon_0$ at $T = 1.5$K of D1 (2+ABC, $\theta = 1.20°$) and D2 (2+ABA, $\theta = 1.19°$), respectively, where $\varepsilon_0$ is the vacuum permittivity. The moiré filling factor $\nu$ is indicated on the top axes. Panel (c) uses a logarithmic color scale. **(d)** $R_{xx}$ along the white dashed trajectories in (b) and (c) at top-gate voltage $V_{tg} = 3.25$ V. The 2+ABA trace is multiplied by 10 for clarity. **(e)** Calculated single-particle band gaps for the two stacking configurations at $\theta = 1.20°$ (top) and longitudinal resistance $R_{xx}$ measured in D1 (2+ABC) and D2 (2+ABA) at $T = 1.5$ K (bottom), plotted as functions of $D/\varepsilon_0$ at full filling ($\nu = +4$). **(f), (g)** Calculated density of states (DOS) at the Fermi energy $E_F$ versus $\nu$ and $D/\varepsilon_0$ for 2+ABC and 2+ABA, respectively, at $\theta = 1.20°$.

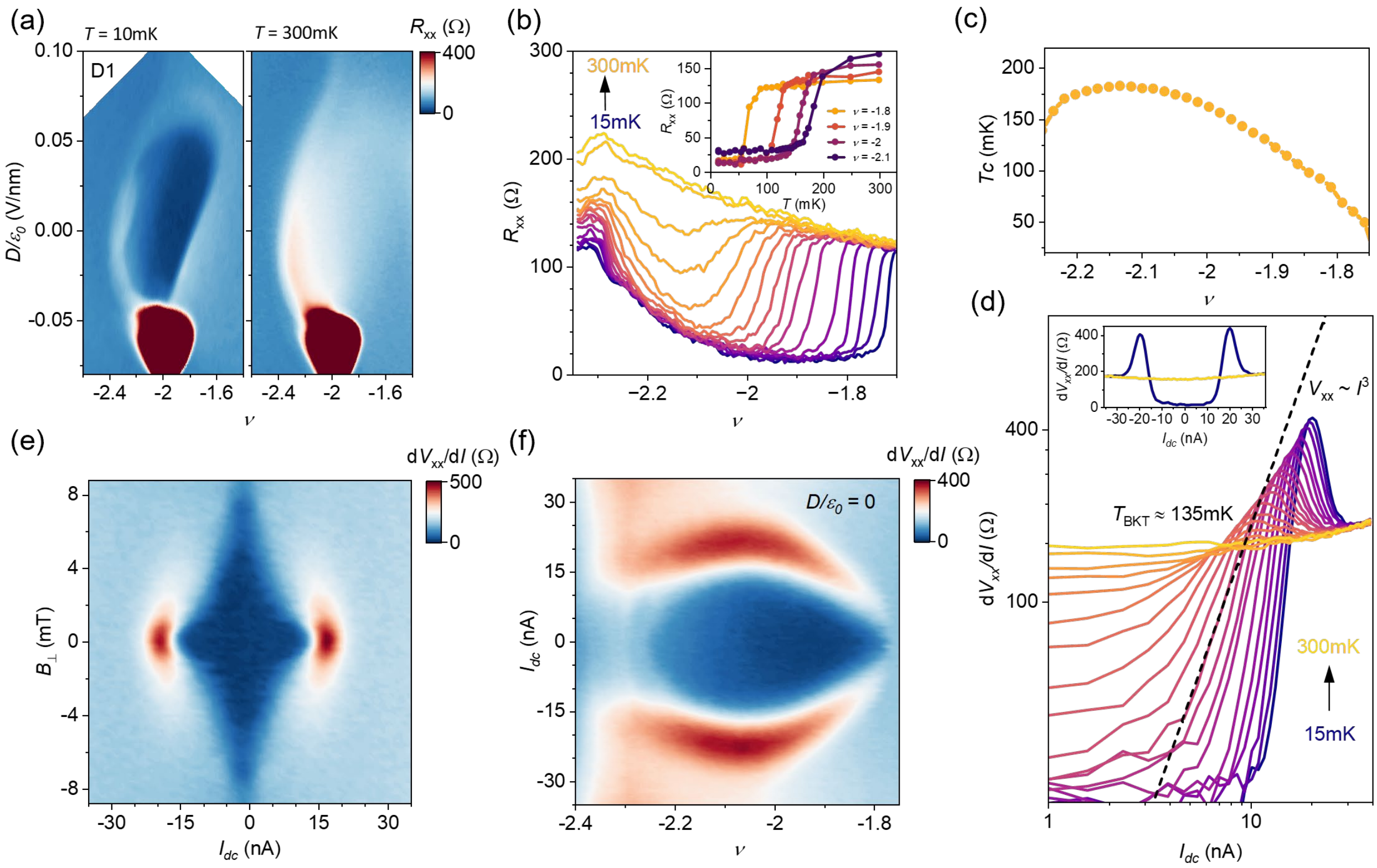

**FIG. 2. Superconducting pocket near $\nu$ = -2 in device D1 (2+ABC, $\theta$ = 1.20°). (a)** Maps of $R_{xx}$ versus $\nu$ and $D/\varepsilon_0$ at $T$ = 10 mK (left) and 300 mK (right). **(b)** $R_{xx}$ versus $\nu$ along $V_{bg}$ = -1 V for $T$ = 15-300 mK. Inset: extracted $R_{xx}(T)$ at selected $\nu$. **(c)** Critical temperatures $T_c$ extracted from (b), defined by $R_{xx}(T_c) = 0.5\ R_N$, where $R_N$ is the normal-state resistance measured above the critical temperature. **(d)** d$V_{xx}$/dI versus $I_{dc}$ on logarithmic axes for $T$ = 15-300 mK. The dashed line indicates d$V_{xx}$/dI $\propto I_{dc}^2$, corresponding to $V_{xx} \propto I_{dc}^3$. The fitted Berezinskii–Kosterlitz–Thouless (BKT) transition temperature is $T_{BKT} \approx$ 135 mK. Inset: d$V_{xx}$/dI versus $I_{dc}$ at $T$ = 15 and 300mK, respectively. **(e)** d$V_{xx}$/d$I$ versus $I_{dc}$ and perpendicular magnetic field $B_\perp$. **(f)** Differential resistance d$V_{xx}$/dI versus $\nu$ and dc bias current $I_{dc}$ at $D/\varepsilon_0$ = 0. Unless otherwise specified, measurements are performed at $T$ = 10 mK.

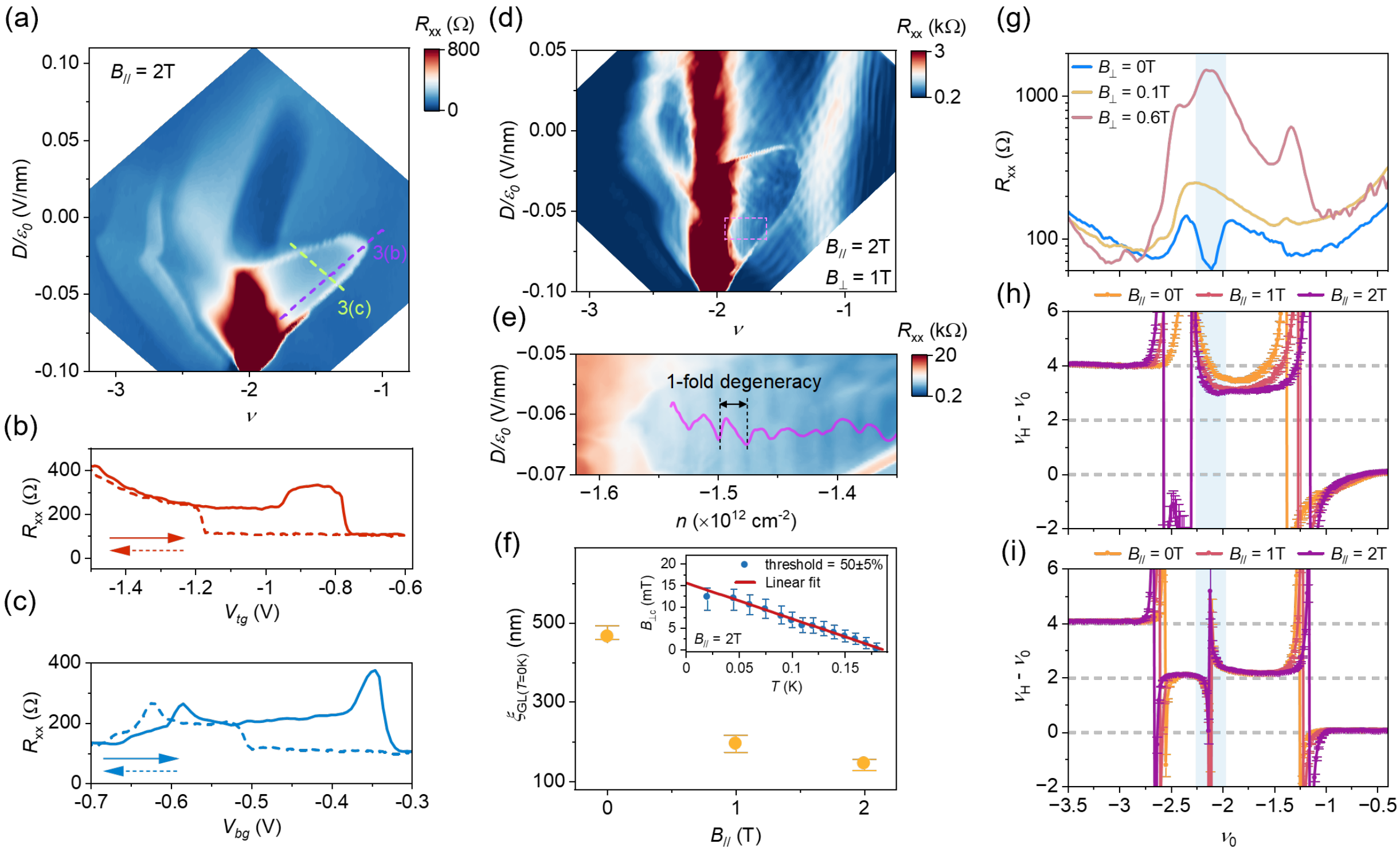

**FIG. 3. Evolution of superconducting and metallic states with in-plane magnetic field in device D1. (a)** Maps of $R_{xx}$ versus $\nu$ and $D/\varepsilon_0$ near $\nu$ = -2 at in-plane magnetic field $B_{//}$ = 2T. **(b), (c)** Hysteresis of $R_{xx}$ near the phase boundary of the quarter-metal (QM) state. The loops are measured by sweeping in the top-gate voltage $V_{tg}$ at fixed $V_{bg}$ = -0.415V and the bottom-gate voltage $V_{bg}$ at fixed $V_{tg}$ = -1.08V, marked by dashed lines in (a). Arrows indicate the sweep directions. **(d)** Maps of $R_{xx}$ versus $\nu$ and $D/\varepsilon_0$ at $B_{//}$ = 2T and $B_\perp$ = 1T. **(e)** Enlarged maps of the boxed region in (d). pink lines show representative oscillation traces after background subtraction, with periods consistent with one-fold degeneracies. **(f)** Coherence length $\xi_{GL(T=0K)}$ versus $B_{//}$. Error bars indicate the range obtained using resistance thresholds of 0.45 $R_n$ and 0.55 $R_n$. Inset: perpendicular critical field $B_{\perp c}$ versus $T$ at $B_{//}$ = 2T, determined using $R_{xx}(B_{\perp c})$ = 0.5 $R_n$. $R_n$ is the normal-state resistance measured above the critical magnetic field. Error bars indicate the range obtained using resistance thresholds of 0.45 $R_n$ and 0.55 $R_n$. The red line represents the linear fit to the Ginzburg–Landau form above 40mK, yielding a zero-temperature coherence length $\xi_{GL(T=0K)} \approx$ 145 nm. **(g)** Symmetrized $R_{xx}$ versus moiré filling factor $\nu_0$ at $B_\perp$ = 0, 0.1 and 0.6T while $B_{//}$ = 0T. Semitransparent blue shading shows the position corresponding to the zero-field SC. **(h), (i)** Antisymmetrized Hall filling $\nu_H$ relative to moiré filling $\nu_0$, $\nu_H$ - $\nu_0$, at $B_{//}$ = 0, 1, 2T, extracted using two different $B_\perp$ fitting windows, [-0.1T, 0.1T] (h) and [0.4T, 0.6T] (i). Hall filling $\nu_H$ is fitted by $\nu_H = (en_0\, dR_{xy}/dB_\perp)^{-1}$. The data were measured at $D/\varepsilon_0$ = 0. Error bars were obtained from one standard error of the slope from the linear fitting.

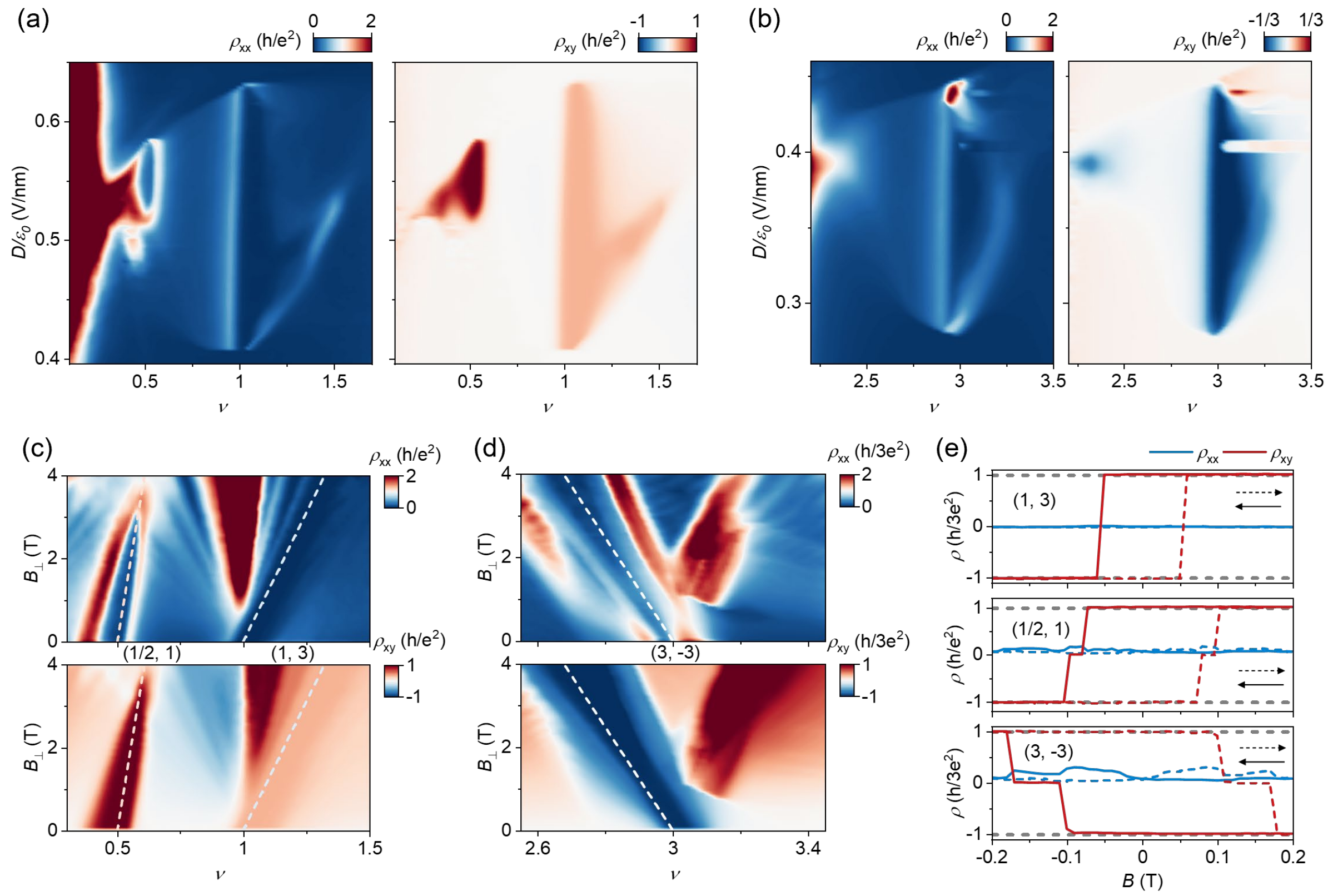

**FIG. 4. Quantum anomalous Hall states on the electron-doped side in device D1.** The longitudinal/Hall resistivity is symmetrized/antisymmetrized, respectively (Method). **(a), (b)** Maps of longitudinal resistivity $\rho_{xx}$ (left) and Hall resistivity $\rho_{xy}$ (right) versus $\nu$ and $D$, measured at $B_\perp = \pm 0.2$T near $\nu = 1/2$ and 1 in (a) and near $\nu = 3$ in (b). **(c), (d)** Landau fan diagrams of $\rho_{xx}$ (top) and $\rho_{xy}$ (bottom) at $D/\varepsilon_0 = 0.554$V/nm near $\nu = 1/2$ and 1 in (c) and at $D/\varepsilon_0 = 0.297$V/nm near $\nu = 3$ in (d). Dashed lines are following the Streda's formula ($\partial n/\partial B_\perp = C \cdot e/h$) and labeled as ($\nu_0$, $C$), where $\nu_0$ is the zero-field moiré filling factor and $C$ is the Chern number. **(e)** Magnetic hysteresis of $\rho_{xx}$ (blue) and $\rho_{xy}$ (red) for the states labeled ($\nu_0$, $C$) = (1, 3), (1/2, 1), and (3, −3), from top to bottom. Arrows indicate the magnetic-field sweep directions.